\documentclass[sigconf]{acmart}
\usepackage[multiple]{footmisc}

\usepackage{times}  % DO NOT CHANGE THIS
\usepackage{helvet} % DO NOT CHANGE THIS
\usepackage{courier}  % DO NOT CHANGE THIS
\usepackage{graphicx}  % DO NOT CHANGE THIS
\usepackage{subfig}
\usepackage{xspace}
\usepackage{balance}
\usepackage{booktabs}
\usepackage{xcolor}% (if used in text)
 \usepackage{blindtext}

\setcopyright{none}
\renewcommand\footnotetextcopyrightpermission[1]{}
\setcopyright{none}
\makeatletter
\renewcommand\@formatdoi[1]{\ignorespaces}
\makeatother

\begin{document}
\title{Under the Spotlight: Web Tracking in Indian Partisan News Websites}

%uncomment 2
\newcommand\sbm[1]{\textbf{\textcolor{brown}{SBM: #1}}	}
\newcommand\nk[1]{\textbf{\textcolor{blue}{NK: #1}}	}
\newcommand\ns[1]{\textbf{\textcolor{red}{NS: #1}}	}
\newcommand\pa[1]{\textbf{\textcolor{purple}{PA: #1}}	}
\newcommand\yv[1]{\textbf{\textcolor{orange}{YV: #1}}	}
\newcommand\va[1]{\textbf{\textcolor{green}{VA: #1}}	}

\author{
Vibhor Agarwal\texorpdfstring{$^{\star} {} ^{\alpha}$}{},
Yash Vekaria\texorpdfstring{$^\star {} ^\alpha$}{},
% Vibhor Agarwal$^\alpha$,
% Yash Vekaria$^\alpha$,
Pushkal Agarwal\texorpdfstring{$^\beta$}{},
Sangeeta Mahapatra\texorpdfstring{$^\gamma$}{},\texorpdfstring{\\}{}
Shounak Set\texorpdfstring{$^\beta$}{},
Sakthi Balan Muthiah\texorpdfstring{$^\alpha$}{},
Nishanth Sastry\texorpdfstring{$^\delta$}{},
Nicolas Kourtellis\texorpdfstring{$^\zeta$}{}
}
\affiliation{
$^\alpha$The LNM Institute of Information Technology, Jaipur, India\\
$^\beta$King's College London, London, United Kingdom\\
$^\gamma$German Institute for Global and Area Studies, Hamburg, Germany\\
$^\delta$University of Surrey, Surrey, United Kingdom\\
$^\zeta$Telefonica Research, Barcelona, Spain\\
$^\alpha$\{vibhor.agarwal.y16, yash.vekaria.y16, sakthi.balan\}@lnmiit.ac.in,
$^\beta$\{pushkal.agarwal, shounak.set\}@kcl.ac.uk\\
$^\gamma$sangeeta.mahapatra@giga-hamburg.de,
$^\delta$n.sastry@surrey.ac.uk,
$^\zeta$nicolas.kourtellis@telefonica.com}

%old authors style
% \author{Vibhor Agarwal}
% \author{Yash Vekaria}
% \affiliation{
%   \institution{The LNM Institute of Information Technology}
%   \city{Jaipur}
%   \country{India}
% }
% \email{vibhor.agarwal.y16@lnmiit.ac.in}
% \email{yash.vekaria.y16@lnmiit.ac.in}

% \author{Pushkal Agarwal}
% \affiliation{
%   \institution{King's College London}
%   \city{London}
%   \country{United Kingdom}}
% \email{pushkal.agarwal@kcl.ac.uk}

% \author{Sangeeta Mahapatra}
% \affiliation{
%   \institution{German Institute for Global and Area Studies}
%   \city{Hamburg}
%   \country{Germany}
% }
% \email{sangeeta.mahapatra@giga-hamburg.de}

% \author{Shounak Set}
% \affiliation{
%   \institution{King's College London}
%   \city{London}
%   \country{United Kingdom}}
% \email{shounak.set@kcl.ac.uk}

% \author{Sakthi Balan Muthiah}
% \affiliation{
%   \institution{The LNM Institute of Information Technology}
%   \city{Jaipur}
%   \country{India}
% }
% \email{sakthi.balan@lnmiit.ac.in}

% \author{Nishanth Sastry}
% \affiliation{
%   \institution{University of Surrey}
%   \city{Surrey}
%   \country{United Kingdom}
% }
% \email{n.sastry@surrey.ac.uk}

% \author{Nicolas Kourtellis}
% \affiliation{
%   \institution{Telefonica Research}
%   \city{Barcelona}
%   \country{Spain}
% }
% \email{nicolas.kourtellis@telefonica.com}

\begin{abstract}
India is experiencing intense political partisanship and sectarian divisions. The paper performs, to the best of our knowledge, the first comprehensive analysis on the Indian online news media with respect to tracking and partisanship.
We build a dataset of 103 online, mostly mainstream news websites.
With the help of two experts, alongside data from the Media Ownership Monitor of the Reporters without Borders, we label these websites according to their partisanship (Left, Right, or Centre).
We study and compare user tracking on these sites with different metrics: numbers of cookies, cookie synchronizations, device fingerprinting, and invisible pixel-based tracking.
We find that Left and Centre websites serve more cookies than Right-leaning websites.
However, through cookie synchronization, more user IDs are synchronized in Left websites than Right or Centre.
Canvas fingerprinting is used similarly by Left and Right, and less by Centre.
Invisible pixel-based tracking is 50\% more intense in Centre-leaning websites than Right, and 25\% more than Left.
Desktop versions of news websites deliver more cookies than their mobile counterparts.
A handful of third-parties are tracking users in most websites in this study.
This paper, by demonstrating intense web tracking, has implications for research on overall privacy of users visiting partisan news websites in India.

\end{abstract}

\maketitle

%uncomment 1: Ref: https://tex.stackexchange.com/questions/826/symbols-instead-of-numbers-as-footnote-markers
%\renewcommand{\thefootnote}{\fnsymbol{$^\star$}}
%\footnotetext[1]{Both the authors contributed equally to this research.}

%new commands
\renewcommand{\thefootnote}{}
\footnote{\texorpdfstring{$^\star$}{}Both the authors contributed equally to this research.}
\renewcommand{\thefootnote}{\arabic{footnote}}
\lhead{Under the Spotlight: Web Tracking in Indian Partisan News Websites}
\rhead{Agarwal V. and Vekaria Y., et al.}

\section{Introduction}
\label{Sec:Intro}

India represents the largest and the most diverse news media market among democracies, with more than 100,000 registered newspapers and 400 news channels\footnote{\scriptsize\url{https://www.indiantelevision.com/regulators/ib-ministry/total-of-television-channels-in-india-rises-to-892-with-three-cleared-in-june-160709}} in 22 scheduled languages.\footnote{\scriptsize{Registrar of Newspapers for India: \url{http://rni.nic.in/}}}
The growth of online news has been the fastest in the emerging markets, with India ranking among the top ten globally when it comes to print and online news media~\cite{GWI2019}. Unfortunately, this growth of online political communications has been accompanied by rising partisanship~\cite{das2020online,mahapatra2019polarisation}. The mainstream news media as major agents of information and influence, become important here.

This paper focuses on major news websites in terms of how they track their users. Tracking allows them to obtain rich information about readers, which may serve their business interest in revenue generation through targeted ads, as well as their political interest in setting agendas. There have been US-based studies about partisan media mostly in terms of their polarizing effects~\cite{Garrett2018,Vargo2019,bhatt2018illuminating} and a few on tracking~\cite{NewsforFree2015,agarwal2020stop}. For India, while there have been a few works on the division in the news media along partisan lines \cite{Partisanship-Freedom}, there is a lack of comprehensive, data-driven research on news websites and tracking behavior.
Indian news media are a major source of information for the population~\cite{Reuters-India-Report}.\footnote{\scriptsize\url{https://bestmediainfo.in/mailer/nl/nl/IRS-2019-Q4-Highlights.pdf}}
Their tracking behavior has socio-political implications as they are, by and large, a trusted source of public information~\cite{TrustinMedia2020}.

In this work, for the first time, we provide a comprehensive study of the news websites in India with respect to partisanship and tracking of online users. We focus on the online platforms of the largest English, Hindi, and regional language news media (including those with print or broadcast platforms and the digital only ones) that can reach more than 77\% of India's population\footnote{\scriptsize{Media Research Users Council: \\\url{https://bestmediainfo.in/mailer/nl/nl/IRS-2019-Q4-Highlights.pdf}}}$^,$\footnote{\scriptsize{Broadcast Audience Research Council, India: \url{https://barcindia.co.in/}}}, making them vulnerable to tracking\footnote{\scriptsize{Personal Data Protection Bill, tabled in Indian Parliament in December 2019, is still with the Joint Parliament Committee for review.}}.
We first identify the major Indian news publications based on their circulation figures from the Registrar of Newspapers for India (RNI) supplemented with Indian Readership Survey of Q4 2019.
We then create a list of 103 news websites, curated primarily from Alexa~\cite{Alexa} and Feedspot~\cite{Feedspot}.
Secondly, with the help of two experts in political science and journalism, alongside data from the Media Ownership Monitor of the Reporters without Borders, which traces associations between the media and political parties and corporate interests~\cite{MediaMonitor}, we label the 103 websites according to their partisanship as Right-, Left-, Centre-leaning, or Unknown (methodology explained in Section~\ref{sec:dataset}).

We address the following questions:
RQ1: What is the extent of tracking across partisan news websites?
RQ2: What kind of tracking methods are used on users?
To answer them, we measure the intensity of user tracking across partisan websites with simple and advanced mechanisms: basic first and third-party cookies, cookie synchronization, device fingerprinting, and invisible pixel-based tracking (Section~\ref{Sec:MeasuringTracking}).

We share our Dataset, OpenWPM Crawls, and Codes publicly with the research community for reproducibility and extension of our work\footnote{\scriptsize{Data and code are available at~\url{http://tiny.cc/india-tracking}}}. From this study, we derive the following key findings (Section~\ref{sec:tracking-results}).
The 103 Indian news websites studied have more than 100K cookies, for an average of over 100 cookies per website, but several websites have much higher number of cookies. For example, $\sim$1400 cookies are set on the first-party -- \textit{Sandesh.com}, by itself and its third-parties. Left- and Centre-leaning websites serve more (median) cookies than Right-leaning websites.
Desktop versions of websites set more cookies than their mobile versions, with interesting exceptions. Third-party domain \textit{doubleclick.net} is present in 86\% of news websites; such ubiquitous presence allows the tracking of a huge proportion of users' browsing histories.

In addition to the large numbers of cookies, we also find evidence of practically every known advanced method of user fingerprinting. Around 18\% of all distinct third-parties, and 25\% of all distinct first-parties in our data are involved in cookie synchronization. Around
50\% of unique user IDs are synced across tracking domains through cookie synchronization.
Cookie synchronization is higher among Left-leaning websites and their third-parties than for Right- and Centre-leaning websites. Over 25\% of news websites use device fingerprinting, which is invisible to the user and invasive to their online privacy. Around
25.7\% of Left, 23.7\% of Right, and 17.9\% of Centre websites employ different fingerprinting scripts to track users.
More than 2.5K invisible (1x1 pixel) images (i.e., 23\% of all sent images) are detected on news website homepages.
Invisible pixel-based tracking is employed more by Centre, followed by Left and then the Right websites.

\section{Background and Related Work}
\label{Sec:Background}

We briefly discuss here the partisan nature of Indian news websites as well as online tracking techniques studied in literature.

\noindent \textbf{Partisan nature of Indian news}:
This paper takes partisanship to mean an adherence to the political beliefs and identification with a political party or cause,
manifesting positively as a civic ideal of shared values or negatively as a pathology where loyalty to a party's ideology/values/goals may trump logic and tolerance to other political views~\cite{MeaningPartisanship2016}.
While numerous political parties exist in India, the three broad strands of political worldviews correspond to three principal political formations at the national level of Indian politics: ``Left'' represented by parties like the Communist Party of India (Marxist),
``Right to Right-Centre'' represented by the Bharatiya Janata Party, and
``Left-Centre'' corresponding to the Indian National Congress. As India is a highly diverse country with their political parties and media reflecting this diversity, we take Right-leaning news media to correspond with the Right to Right of the Centre spectrum of ideologies, the Left-leaning news media to correspond with the Left to Left of the Centre spectrum, and the Centre-leaning media to be positioned in between the Right-Centre and the Left-Centre.
The growth of heightened political partisanship may have a dramatic impact on media behavior and their influence on public opinion, especially if they intensely track users.

\textbf{Online tracking ecosystem and measurements}:
With the rise of online information consumption, online platforms have attracted third parties for online advertising~\cite{mccoy2007effects,papadopoulos2017if}.
These advertisements are strategically drafted and placed on websites to get more user attention including pop-ups and banners~\cite{mccoy2007effects,speicher2018potential}.
These websites track users by injecting cookies at the users' side~\cite{englehardt2015cookies,binns2018measuring,vallina2016tracking,hu2020websci} for content personalization and improving user experience.
However, cookies and other data are also shared with other third parties, raising privacy concerns.
Users have an option to accept or reject these third-party cookies, but many users are not aware of the consequences if they accept them.
These websites also use more sophisticated tracking techniques like cookie synchronization~\cite{acar2014web,englehardt2016online,papadopoulos2019cookie,agarwal2020stop,urban2020measuring,hu2020websci}, device fingerprinting~\cite{mowery2012pixel,englehardt2016online}, and invisible (1x1) pixel-based tracking~\cite{fouad2018missed}.
Since users are often unaware of their presence, such methods pose a greater privacy threat to the websites' visitors.
Studies have shown that some popular trackers like Doubleclick and Google Analytics (both Google trackers) can be present in up to 50\% and 70\%, respectively, of top one million visited websites~\cite{englehardt2016online}.
Specifically, news websites have seen large volume of trackers and advertisements including political campaigns~\cite{englehardt2016online,agarwal2020stop,papathanassopoulos2013online}.
Among USA news websites, Right-leaning websites track users more and have high cookie synchronization within the partisan group websites~\cite{agarwal2020stop}.
Having said that, less is known about the tracking ecosystem of Indian news media, which has recently seen exponential growth in online consumption.
There are studies in online engagement (including social media) showing polarization and media bias, but none covers the exposure of user data to the tracking world~\cite{mahapatra2019polarisation,10.1145/3209811.3209825,qayyum2018exploring}.
With our work, we aim to fill this gap by measuring the extent to which users are exposed to a high amount of web tracking, using the aforementioned four tracking techniques.
We also explore tracking on desktop and mobile platforms in Indian news media with partisan leanings.

\section{Data Collection and Labeling}
\label{sec:dataset}

\begin{figure*}
\centering
  \includegraphics[width=\textwidth]{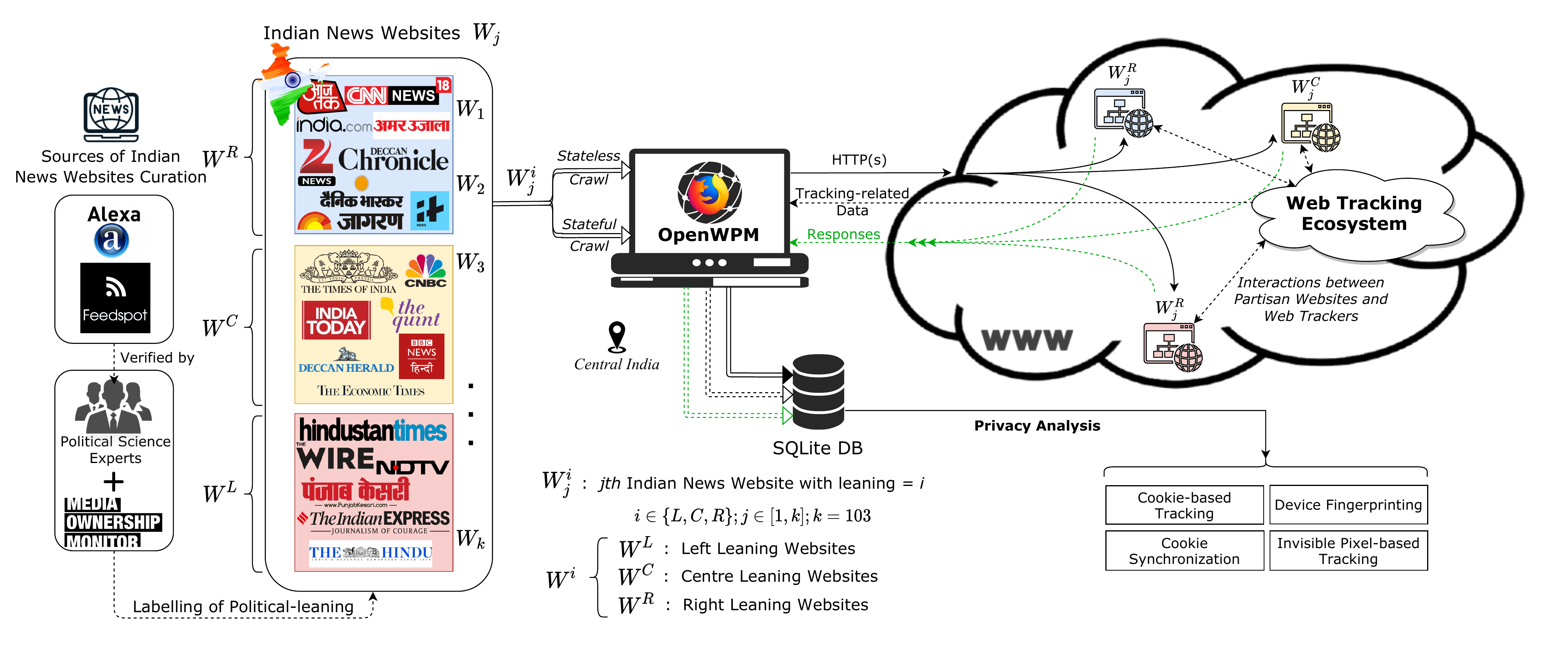}
  \caption{Our framework for labeling Indian news websites along partisan lines and collecting web traffic data for studying web tracking mechanisms. Colors represent party-leaning: Right=Blue, Centre=Yellow, and Left=Red.}
  \label{fig:Model}
\end{figure*}

Here, we discuss the methodology followed to curate a list of top news websites in India, including metadata crawled for each using \emph{Feedspot}~\cite{Feedspot} and \emph{Alexa.com}~\cite{Alexa}, to label these websites based on their political leanings (Sec.~\ref{subsec:website-labeling}).
Furthermore, in Sec.~\ref{subsec:dataTraffic-collection}, we provide details of our website traffic crawling using \emph{OpenWPM}~\cite{englehardt2016online,openwpm-code}, a tool for desktop browser automation and crawling, and \emph{Cookies.txt}~\cite{cookiesChromeExt}, a browser plug-in for mobile browser automation.

\subsection{Websites Partisan Labeling}
\label{subsec:website-labeling}

We follow the methodology outlined in Figure~\ref{fig:Model} (left part) for website list creation and partisanship labeling.

\noindent\textbf{List Creation:}
We first examined a list of 141 top Indian news websites on the Web (ranked as on 28 April 2020) provided by Feedspot~\cite{Feedspot}.
This website, maintained by over 25 experts, is updated daily and covers a wide range of factors to rank and discover the most prominent online news websites in India.
They curate websites whose publishers explicitly publish their content via Feedspot, as well as by monitoring search engines and social media through in-house media tools.
The next list of websites we studied is from Alexa (29 April 2020)~\cite{AlexaTop49News}.
Alexa Internet, Inc., is an American Web traffic analysis company, whose toolbar gathers information of around 30 million websites across the globe, based on their internet browsing behavior and traffic patterns.
Their website stores the data and provides extensive analysis of the websites.
From Alexa, we got a list of 49 top Indian news websites based on their online popularity and traffic. Some of them were common with the Feedspot data. We combined Feedspot and Alexa lists to obtain a list of 153 websites.

A large portion of news consumption in India happens through online platforms (Facebook, Twitter, and Instagram) rather than TV/Radio~\cite{Reuters-India-Report}.
Therefore, we further augment our data by visiting each website's Facebook, Twitter, and Instagram pages for metadata collection.
After opening a particular website on Facebook, Twitter or Instagram, we performed (in April 2020) a breadth first search on other `Indian news page recommendations' shown in the right-side panel under the heading of ``Related Pages'' in Facebook, ``You might like'' in Twitter, and ``Related Accounts'' at the bottom in Instagram. We added to our list all Indian news media shown in recommendations (as described above) while visiting the social media pages of initially curated websites. In the second-iteration, we repeated this with newly collected news media from the first-iteration. We repeated this approach up to five times, by which we observed that 90\% of  recommendations were already in our dataset. Using this approach, we added to our list 65 new Indian news media leading to a total of 218 websites.
Then we removed websites with inactive web pages and retained only those which had more than 10K followers on at least one of the three social media platforms investigated (to ensure we only include the popular ones).
Our final list has 123 Indian news websites, spanning nine languages and 28 states. All have an online website, which can be freely accessed over the internet.
Out of 123 websites, 10.56\% are popular as TV channels, 53.66\% are print media and remaining 35.78\% only have a website (no TV channel or print media).
We determine popularity in terms of viewership/readership in TV/print media.

\noindent\textbf{Website Labeling:}
In order to understand and categorize websites based on their partisan leanings, we undertook a three-step labeling process.
First, we approached two political science and journalism experts who manually coded the political leanings of these websites.
This approach has been used by media monitors at Buzzfeed News\footnote{\scriptsize{https://www.buzzfeed.com/craigsilverman/inside-the-partisan-fight-for-your-news-feed}} in past studies to review political leaning in the US news ecosystem.
Second, we checked for their partisan associations from Media Ownership Monitor~\cite{MediaMonitor} including data on parent company.
The labeling was then done along a spectrum of Right (Conservative: Right to Right-Centre), Left (Liberal: Left to Left-Centre), and Centre (i.e., less biased or a combination of both Left and Right, that is, when the same parent company has two ideologically different news sites) categories based on ownership and ideological association.
20 websites were discarded due to uncertainty in their leaning. And the remaining 103 websites were labeled with a partisan leaning and considered for our study.
The inter-annotator agreement between experts, measured by Cohen's Kappa, is 0.97. Throughout the paper, we use this categorization, with short names: ``Left'' for ``Left to Left-Centre'', ``Right'' for ``Right to Right-Centre'', and ``Centre'' for ``Centrist or representing view-points of Right and Left''. Our dataset consists of \textit{40} Left-, \textit{26} Centre-, and \textit{37} Right-leaning websites.

\subsection{Websites Traffic Data}
\label{subsec:dataTraffic-collection}

We start our data collection using OpenWPM~\cite{englehardt2016online} by performing five stateless crawls, while visiting the websites' homepages from Central India between August 10, 2020 to August 30, 2020.
Stateless crawls make each website visit independent.
Parallel browser instances were launched to allow multiple, simultaneous crawls of these news websites from a single location.
We performed such crawls across different times and days to account for infrequent but unavoidable network errors during each crawl. We recorded more than 100K cookies in total.

We also performed five time-variant and order-variant, stateful crawls of the websites' homepages from September 01, 2020 to September 15, 2020.
Stateful crawls are important since we want to study tracking mechanisms such as cookie synchronization (CS).
CS requires state information to be maintained across different websites and visits, to detect if user IDs from previous visits are being synced in future visits and with other websites and their third-parties.
Time-variance is applied by crawling on different days with days-long time between crawls.

Order-variant means the websites are visited in a shuffled order for each crawl, for the results to be independent of the website ordering. In stateful crawls, no parallel browser instances are launched to detect third-parties that indulge in cross-site tracking of users.

For 23 of the 103 websites, we also find manually that they serve separate mobile versions.
Therefore, we perform five additional crawls for these mobile websites to compare tracking behavior in desktop websites and their mobile counterparts.
The crawling for mobile websites uses \emph{Cookies.txt}, a Firefox Plug-In ~\cite{cookiesChromeExt} to get browser cookies information.
We automate this process using Selenium\footnote{\scriptsize{\url{https://www.selenium.dev/documentation/en/}}}.
At first, a Firefox browser is set to not block any type of cookies. Further steps include opening a Firefox Mobile Emulator in an incognito mode, loading the plug-in, visiting the mobile versions of the websites' homepages (e.g., \emph{m.timesofindia.com}), and storing cookies information.
In these five crawls, we store 1400 cookies in total.

\section{Measuring Tracking Mechanisms}
\label{Sec:MeasuringTracking}
In this section, we detail the methodology to measure various tracking methods used by Indian news websites and the associated ad-ecosystem --
Figure~\ref{fig:Model} (right part).

\subsection{First and Third-party Cookie Analysis}
\label{subsec:cookie-analysis}

To perform the cookie-based analysis, we use the \emph{javascript\_cookies} table of SQLite dump from the OpenWPM crawled data.
This data provides information on all different types of cookies being set by different domains.
In addition, we use the Disconnect List\footnote{\scriptsize{https://github.com/disconnectme/disconnect-tracking-protection}}, which is extensively used by the research community to report known tracking domains, and categorize them into eight distinct categories: Advertising, Analytics, Content, Social, Fingerprinting, Cryptomining, Disconnect, and Unknown.
We use this list to understand the distribution of cookies across these categories.

\subsection{Cookie Synchronization Analysis}
\label{subsec:csync-analysis}

Cookie synchronization (CS) is a cross-site tracking mechanism that enables two trackers to generate a detailed browsing profile of the user, by sharing unique user IDs with each other.
CS circumvents the Same-Origin Policy (SOP)\footnote{\scriptsize{SOP allows tracking domains to access only cookies set by them.}}.
Past works have studied CS in different contexts~\cite{acar2014web,falahrastegar2016tracking,englehardt2016online,papadopoulos2019cookie,agarwal2020stop,urban2020measuring,hu2020websci}).
However, CS has never been studied specifically for Indian news websites along partisan lines or with respect to the privacy implications that it has in the context of India.
CS can be abstracted as a two-step process.
In the first step, a unique user ID is exchanged between two TPs in the form of HTTP(s) requests, responses, or redirects in an effort to learn the identity of the given user on the web.
This ID can be used to aggregate user information by a variety of means~\cite{gonzalez2017cookie} through step two.
In the second step, domains exchange or merge the identified user's data including browsing histories, browsing patterns, and interests through a separate ``data sharing channel'' to build a complete, consolidated user profile.

\noindent\textbf{Privacy impact:}
Tracking and targeting based on CS primarily helps advertisers~\cite{lerner2016internet}, especially in programmatic (real-time bidding) advertising, where data sharing and purchasing involves CS for better targeting~\cite{ghosh2015match}.
As a result of CS, trackers are able to track a given user over a larger set of websites, where they may not even be embeded as TPs.
In fact, repetitive CS across websites can enrich a particular user’s profile built by trackers, helping them to precisely track and target a user over time.
Also, server-to-server exchanges of user data (CS step 2 above) have become common~\cite{englehardt2016online}, enabling deeper user profiling.

\noindent\textbf{Methodology:}
We capture CS for websites in our dataset using similar methodology of past studies~\cite{acar2014web,falahrastegar2016tracking,papadopoulos2019cookie}.
We use the fundamental structure of the open-source python code from~\cite{acar2014web} (referred to as \texttt{CSCode} hereafter) and make modifications to work for our scenario: unlike~\cite{acar2014web} that crawled data simultaneously on two machines before analyzing them with \texttt{CSCode}, we perform time-variant crawls (Sec.~\ref{subsec:dataTraffic-collection}).

For each crawl, we detect CS for each leaning group and a combination of them.
For example, while studying CS between Left and Right, we iterate over all distinct pairs of websites \texttt{(w1,w2)} where \texttt{w1} is any website which is Left only, while \texttt{w2} is Right only (with \texttt{w1!=w2} and \texttt{(w1,w2)} ${\displaystyle \equiv}$ \texttt{(w2,w1)}).
Since we have 39 Left and 37 Right websites, there are 39x37=1443 total pairs.
For intra-party comparisons like Right-Right for instance, the total unique pairs will be computed as ${}^{37}C_{2}$ = $666$.
Next, for each pair, we consider all the HTTP(s) request, response, and cookies data related to \texttt{w1} and \texttt{w2}, and use \texttt{CSCode} to search for IDs synced between FPs and TPs while visiting \texttt{w1} and \texttt{w2}.
We try all possible combinations of website pairs falling into different partisan lines, i.e.:

\begin{itemize}
\setlength\itemsep{-0.4em}
\item $w1\in W^L\ \mbox{and}\, w2\in W^L$ ; $w1\in W^R\ \mbox{and}\, w2\in W^R$
\item $w1\in W^C\ \mbox{and}\, w2\in W^C$; $w1\in W^L\ \mbox{and}\, w2\in W^R$
\item $w1\in W^L\ \mbox{and}\, w2\in W^C$; $w1\in W^R\ \mbox{and}\, w2\in W^C$
\end{itemize}
Since~\cite{acar2014web} is an older paper on CS, we validated \texttt{CSCode}, as well as various parameters used with recent works on CS~\cite{papadopoulos2019cookie,agarwal2020stop,urban2020measuring,hu2020websci}).
We made the following key changes to ensure result correctness.
First, for each URL, \texttt{CSCode} extracts the top-level-domain (e.g., \textit{com} from \textit{rtb.gumgum.com}) in~\cite{acar2014web}. However, it is not relevant to study CS across such top-level domains. Instead, we follow~\cite{papadopoulos2019cookie} and map all domains (from cookies, requests, response URLs, etc.) to the high-level domains returned by the WhoIS tool\footnote{\scriptsize{https://www.whois.com/}} (e.g., \textit{rtb.gumgum.com} is mapped to \textit{gumgum.com} as obtained from WhoIS).
Second, \texttt{CSCode} constraints minimum length of an ID to be \texttt{6} characters. However,~\cite{urban2020measuring} suggests to discard shorter IDs, since they do not contain sufficient entropy to represent a user ID. We follow~\cite{papadopoulos2019cookie} and use threshold of \texttt{11} characters to minimize false positives. Interestingly, the shortest ID detected in our data is \texttt{12} characters long.
Third, we upgraded \texttt{CSCode} to support python3 and dependencies.

\noindent \textbf{Limitations:} \texttt{CSCode} gives a strict conservative ID detection with fewer false positives~\cite{acar2014web}.
However, false negatives may occur when ID is shared in URL parameters in an encoded or encrypted format~\cite{papadopoulos2019cookie,bielova2020missed}, or when ID strings are hidden inside the longer strings with non-standard delimiters.
According to~\cite{acar2014web}, the adversarial trackers could have short-lived cookies\footnote{\scriptsize{As~\cite{acar2014web}, we consider cookies with expiration date $\leq$ 30 days}} mapped to user IDs at the backend-server to later on track the user.
Such cases are not captured by our code.
Hence, our results represent a \textit{lower bound} on the actual CS taking place in a real-time scenario.

\subsection{Device Fingerprinting Analysis}
\label{subsec:fingerprinting-analysis}

\noindent \textbf{Privacy impact:} A device or browser fingerprinting is a powerful technique that websites and TPs use to identify unique users and track their online behavior.
This method collects information about the user's browser type and version, operating system, time-zone, language, screen resolution, and other settings.
It can lead to serious privacy issues as users are oblivious to this happening, and can have important implications on the way third-parties track users across the Web \emph{without cookies} in the future.

\noindent \textbf{Methodology:} Our fingerprinting measurement methodology~\cite{englehardt2016online} utilizes data collected by OpenWPM, as described in Sec.~\ref{subsec:dataTraffic-collection}.
In particular, we detect different types of fingerprinting such as canvas, WebRTC, and audioContext, by checking webpages and the interfaces they call, such as \textit{HTMLCanvasElement} and \textit{CanvasRenderingContext2D} for canvas, \textit{RTCPeerConnection}, \textit{createDataChannel} and \textit{createOffer} for WebRTC, and \textit{AudioContext} and \textit{OscillatorNode} for audioContext.

\subsection{Invisible Pixel-based Tracking Analysis}
\label{subsec:pixel-analysis}

\noindent \textbf{Privacy impact:} Invisible pixels are 1x1 pixel images that do not add any content to the websites hosting them.
TPs use these invisible pixels to track user's behavior on a website.
Whenever a website loads, it sends subsequent requests to the server to load various assets like images, ads, and other media on the website.
To load these invisible (1x1) pixels on the websites, TPs send some information using the requests sent to retrieve the images.
Crucially, the users are unaware of the pixels' existence on the websites and that these pixels report user's activity.
Therefore, every such pixel represents a threat to the user's privacy.

\noindent \textbf{Methodology:}
We follow~\cite{fouad2018missed}, and for every crawl using OpenWPM, we store all HTTP requests, responses, and redirects, along with response headers, to capture the communication between a client and a server.
We then filter HTTP requests and responses by checking the \textit{content-type} in the response header.
If the \textit{content-type} is an \textit{image}, the corresponding requests and responses are for images.
Next, we check for \textit{content-length} in the response headers to filter out only those HTTP requests and responses with \textit{content-length} less than 1KB.
This threshold is used to save storage space (i.e., not to store all images but only probable 1x1 pixel images).
In~\cite{fouad2018missed}, they use 100KB threshold, but this is a very large size for such 1x1 pixel images.
In fact, we found all detected invisible pixels in our dataset are less than 1KB in size.
All such images are downloaded using the image's URL recorded in the filtered HTTP requests and responses and then checked for the image's dimensions.
If both height and width of an image are 1 pixel, then the image is labeled as invisible pixel.
The corresponding HTTP request/response, image URL, content length, and third-party setting of each invisible pixel are recorded for further analysis.

\section{User Tracking vs. Partisanship}
\label{sec:tracking-results}

In this section, we present our privacy analysis on the partisan websites of our dataset, and how they track users.
We start with cookie-based tracking analysis (Sec.~\ref{subsec:cookie-results}).
We then study more complex tracking techniques such as cookie synchronization (Sec.~\ref{subsec:csync-results}), device fingerprinting (Sec.~\ref{subsec:fingerprinting-results}), and invisible pixel-based tracking (Sec.~\ref{subsec:pixel-results}).

\begin{figure}
\centering
  \includegraphics[width=\columnwidth, height=5cm]{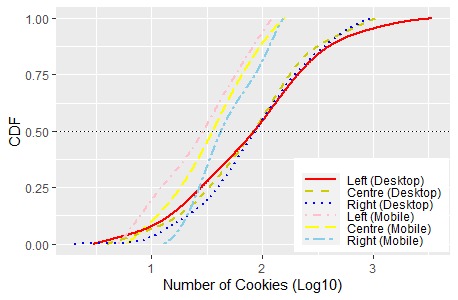}
  \caption{CDF of number of cookies for Left, Centre, and Right-leaning news websites, for their desktop and mobile versions (if available).}
  \label{fig:cookiesCDF}
\end{figure}

\subsection{Number of cookies}
\label{subsec:cookie-results}

We analyze 100K cookies placed by FPs and TPs while visiting the 103 Indian news websites.
Figure~\ref{fig:cookiesCDF} shows the CDF of the number of cookies for all the Left-, Centre-, and Right-leaning news websites available for desktop (103) and mobile (23) versions of the websites.
The median number of cookies are 86, 84, and 92 for Left-, Right-, and Centre-leaning desktop websites, and 30, 42, and 36, respectively, for mobile websites.
Therefore, in all political leanings, websites for desktop push more cookies to the user's browser than mobile versions (in median).
In mobile versions, Centre and Right websites track users more compared to the Left by 1.2 and 1.4 times (KS-value: 0.33, p-value: 0.007), respectively, and Right websites tracks more than Centre websites by 1.2 times (KS-value: 0.28, p-value: 0.054).
In desktop versions, median numbers are close for all leanings.
The Right websites have fewer cookies than the Left, and the Left has fewer than the Centre.
Interestingly, when considering the case of websites for desktop delivering a lot more cookies than the median, Left tracks more than the Right and Centre.
For example, \textit{sandesh.com}, which is in the Left to Left-Centre political spectrum, has the highest number of cookies: more than 1400 cookies (median over five crawls). These cookies are set by the FP and TPs on this website. When desktop websites have cookies less than the median, the trend is reversed, i.e., Right tracks more than Left and Centre.

\begin{figure}
\centering
  \includegraphics[width=\columnwidth]{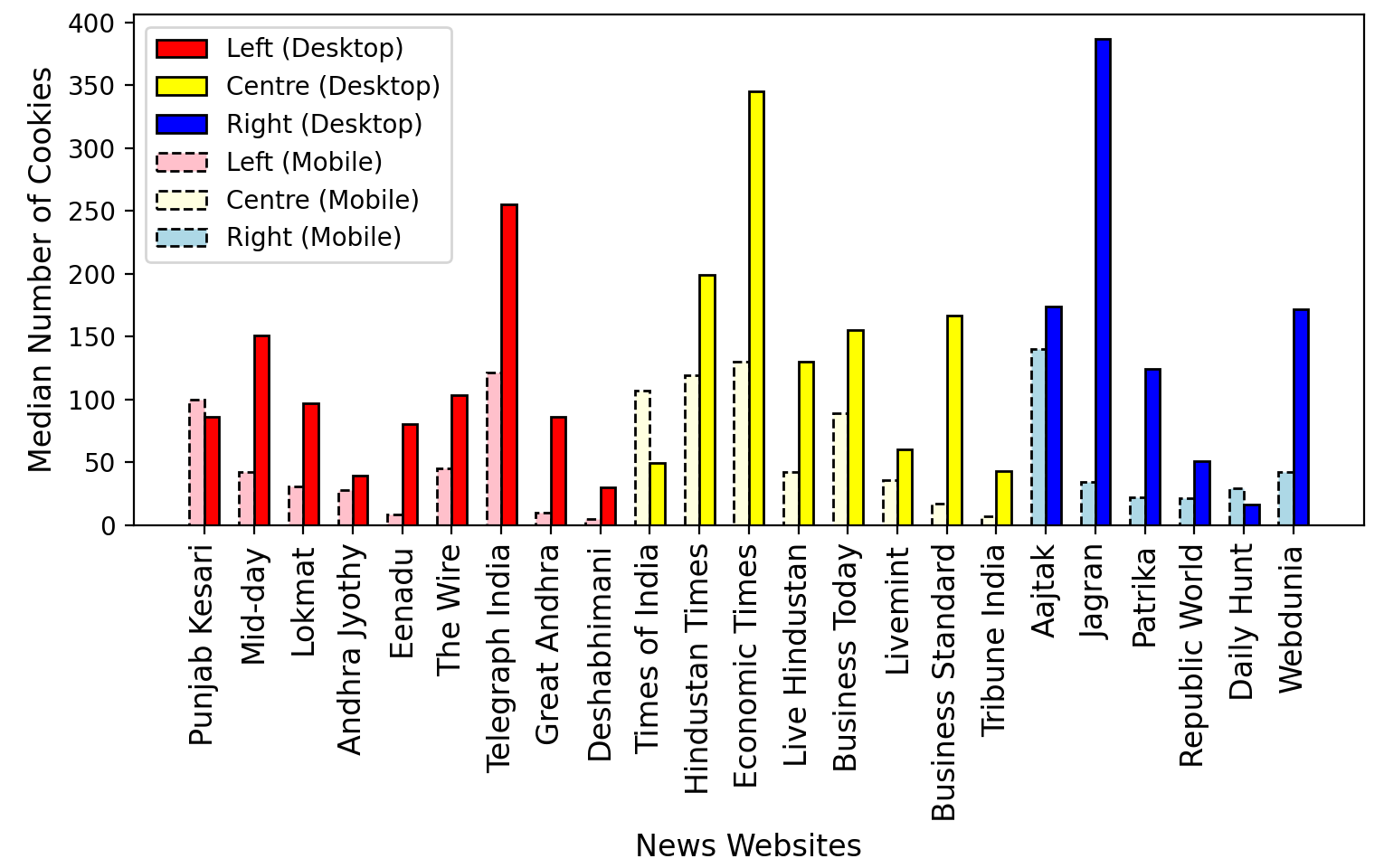}
  \caption{Median number of cookies in mobile vs. desktop versions for 23 news websites, grouped by political leaning in decreasing order of their Facebook followers.}
  \label{fig:mobileCookies}
\end{figure}

The different versions for desktop and mobile platforms for the same news website imply opportunity for collaboration or data leakage between the two tracking ecosystems across different devices.
In Figure~\ref{fig:mobileCookies}, we compare the total number of cookies for each of the 23 news websites with mobile and desktop versions.
Most websites (20/23) set more cookies in their desktop as compared to their mobile versions.
Interesting exceptions are \textit{Times of India}, \textit{Punjab Kesari}, and \textit{Daily Hunt}, which set more cookies in their mobile websites.
More cookies indicate higher intensity of tracking as well as network activity (for storing, updating, and synchronizing said cookies) between the browser and server.
Therefore, such (mobile) websites neither respect users' privacy nor consider the mobile device's limited resources regarding power and bandwidth (data) consumption.

We further investigate the difference in tracking between mobile and desktop, and study the unique TP domains that are present in mobile, desktop, or both versions.
On one hand, we find 68\% of TPs exist in both mobile and desktop versions, allowing them to perform in-depth monitoring of (same) users, and linking them across multiple devices.
On the other hand, we find 16\% of TPs exist only on mobile versions.
For e.g., websites such as \textit{Times of India} and \textit{Punjab Kesari} have more than 50\% of their TPs present in their mobile versions and not in their desktop versions.

\begin{figure}
\centering
  \includegraphics[width=\columnwidth, height=5cm]{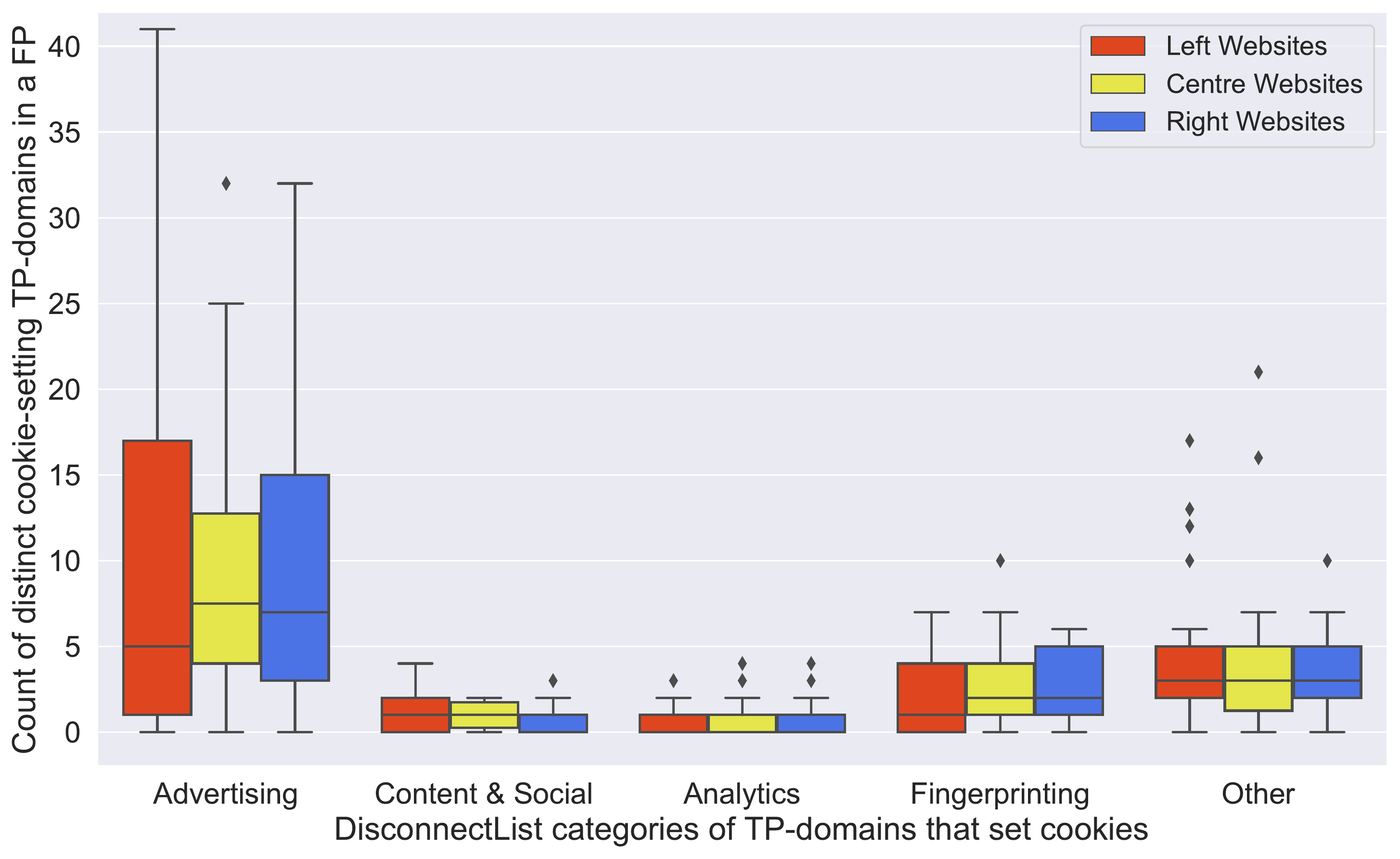}
  \caption{For each first-party (FP), the distribution of count of distinct cookie-setting third-parties (TPs) by DisconnectList categories.}
  \label{fig:DL_boxplot}
\end{figure}

We also study the type of TPs that set cookies on browsers, using the Disconnect List (DL).
Note: we group together ``Cryptomining'', ``Disconnect'' \& ``Unknown'' as ``Other''.
Figure~\ref{fig:DL_boxplot} shows the box-plot distribution of each category.
Statistically, with a KS-value 0.35 (p-value: 0.0195), the largest portion of TP domains is advertising and observed across all partisan websites, with Centre and then, Right being the most frequent.
This is unsurprising since most news websites are funded by display ads.
Interestingly, the second most frequent category (apart from ``Other'') is TP domains performing fingerprinting (KS-value: 0.31, p-value: 0.0534). When compared with medians, we again observe Centre and Right websites being more intense with fingerprinting than Left.
We investigate such domains further in Sec.~\ref{subsec:fingerprinting-results}.

\begin{figure}
\centering
  \includegraphics[width=\columnwidth]{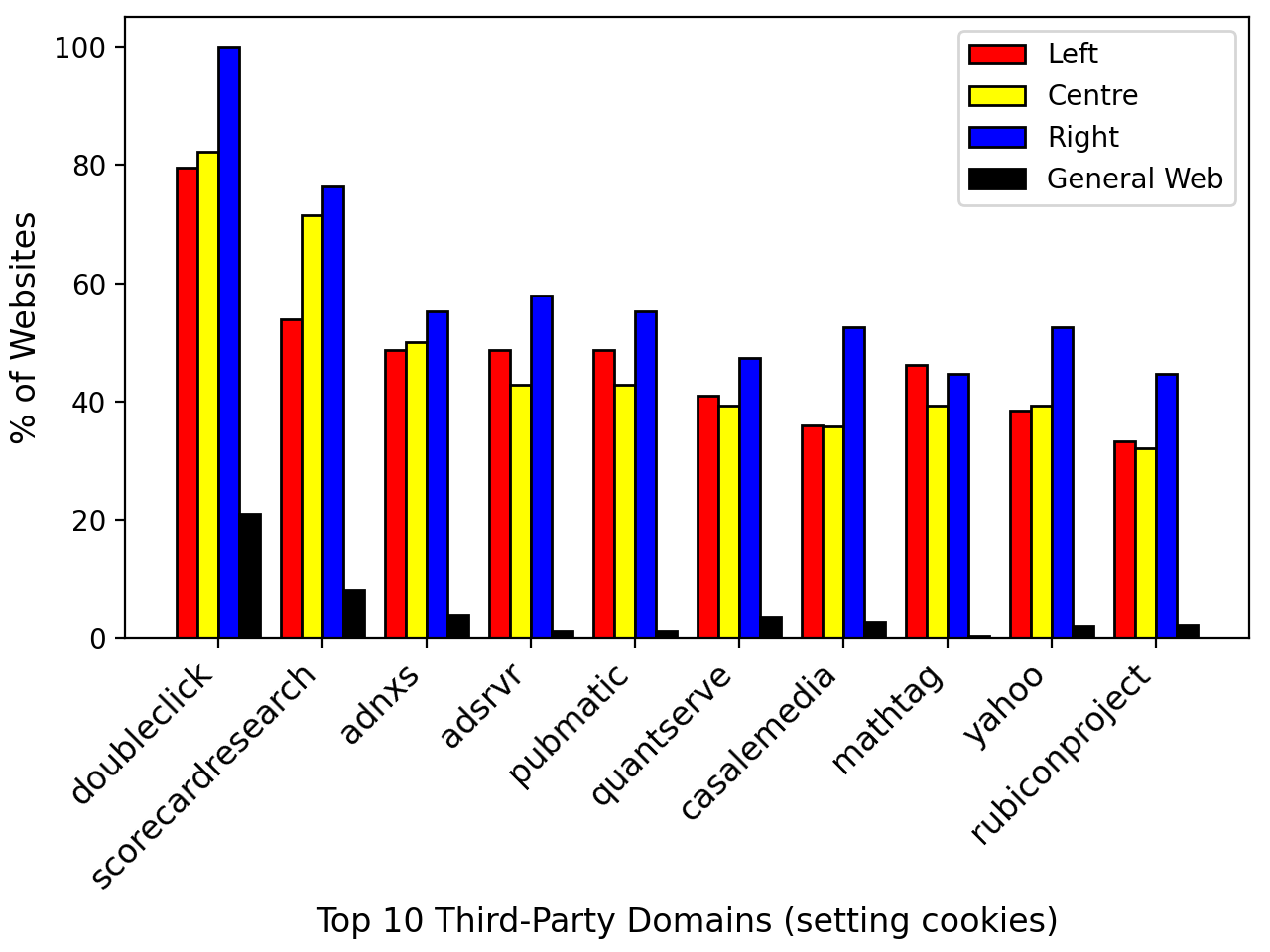}
  \caption{Top 10 TP domains setting cookies in Left, Centre, or Right-leaning news websites.
  Their presence on general web is also plotted for comparison.}
  \label{fig:topDomains}
\end{figure}

Finally, we look into the top TP domains involved in cookie-based tracking.
Figure~\ref{fig:topDomains} shows the top 10 TPs, per political leaning of the first-party website embedding them. We also compare the embeddedness of these TPs with their appearance in the ``general web''. This is to understand how much more or less intensely these TPs track users visiting Indian news websites compared to the general web, following the same strategy as in \cite{agarwal2020stop}. For general web, we crawl data from \textit{whotracks.me}, the percentage of websites in which detected third-parties embed their cookies on the Web. We find these TPs are more embedded in the Right-leaning websites than Left or Centre. Unsurprisingly, \textit{doubleclick.net} is present in most websites in our list: 100\% of Right, 80\% of Left, and 82\% of Centre websites, while in general web, it is tracking only 21\% of websites. Additionally, we look at the portion of cookies contributed by these TPs. We find \textit{pubmatic.com} sets most cookies, contributing an overall 9\% of cookies in our data. Also, the top 10 (2\%) TPs set 42\% cookies in our dataset.

\noindent \textbf{Takeaways:}
Desktop versions of websites set more cookies than mobile.
Also, Right- and Centre-leaning websites embed more Advertising and Fingerprinting TPs than Left-leaning websites, including the top entity \textit{doubleclick.net}.
In general, a handful of TPs provide high coverage of users across all political spectrum of Indian news websites.

\subsection{Cookie Synchronization}
\label{subsec:csync-results}

\begin{table}
    \centering
    \caption{
    Statistics on cookie synchronizations detected between first party (FP) and third party (TP), or TP-TP domains, for all combinations of FP website pairs crawled,
    e.g., ``Right-Left'' means first a visit to a Right-leaning website and then a visit to a Left-leaning website (or vice-versa).
    }
    \scriptsize
    \begin{tabular}{|c|c|c|c|}
    \hline
    \multicolumn{1}{|c|}{\textbf{Leaning}} &
    \multicolumn{1}{|c|}{\textbf{Avg. ID syncs}} &
    \multicolumn{1}{|c|}{\textbf{Avg. ID syncs}} &
    \multicolumn{1}{|c|}{\textbf{Avg. ID syncs}} \\
    \textbf{Group} &
    \multicolumn{1}{c|}{\textbf{per unique ID}} &
    \multicolumn{1}{|c|}{\textbf{per TP-TP pair}} &
    \multicolumn{1}{|c|}{\textbf{per FP-TP pair}} \\
    \hline
    %\midrule
    Right-Right     & 2.59 & 3.83 & 1.65 \\
    Left-Left       & 4.67 & 4.45 & 2.23 \\
    Centre-Centre   & 3.37 & 3.00 & 1.71 \\
    Right-Left      & 4.75 & 4.06 & 1.45 \\
    Right-Centre    & 3.45 & 3.45 & 1.63 \\
    Left-Centre     & \textbf{5.92} & \textbf{4.81} & \textbf{2.46} \\
    \hline
    %\bottomrule
    \end{tabular}
    \label{tab:CSresults}
\end{table}

We compute cookie synchronization (CS) for all stateful crawls as described in Sec.~\ref{subsec:csync-analysis}, and summarize results across different partisan leaning groups, as shown in Table~\ref{tab:CSresults}.

In general, we see that any user browsing that involves visiting a Left-leaning website (before or after a Left, Right or Centre website) leads to an elevated number of CSs per unique ID, in comparison to only Right- or Centre-leaning websites (first column of Table~\ref{tab:CSresults}).
This is also the case for CSs detected between TP-TP pairs.
TPs in Centre-Centre group seem to perform the least amount of such CSs in comparison to other groups.
Finally, Left-Left and Left-Centre have the highest CSs in FP-TP pairs in comparison to other groups.
Right-related groups perform the least CSs.
% This is contrasting to the number of cookies set since highest cookies are set in Right websites.

\begin{figure}
\centering
  \includegraphics[width=\columnwidth, height=5cm]{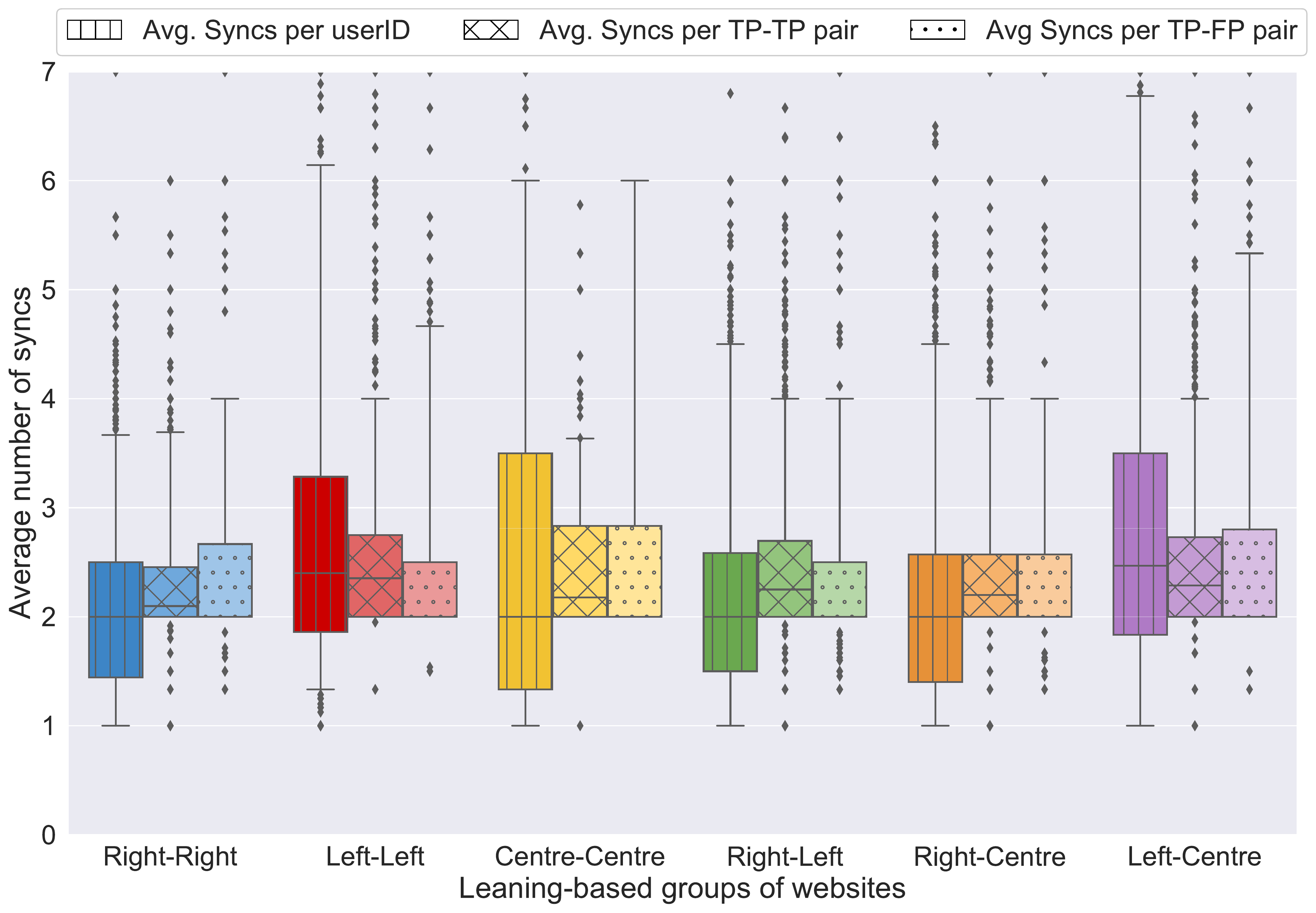}
  \caption{Distributions of average number of CSs per ID, with respect to political leaning groups and combinations.}
  \label{fig:IDs-in-Sync}
\end{figure}

In Figure~\ref{fig:IDs-in-Sync} we look at the distribution of CSs performed per pair of websites visited, per combination of partisan website groups.
With a KS-value of 0.0748 at 0.0029 significance, the highest number of CS happens when Left-Left (i.e., intra-partisan) group of websites is visited. Similarly, among the inter-partisan groups, Left-Centre website visits involve high CS tracking (KS-test: 0.0431, p-value: 0.0003)

\begin{figure}
\centering
\includegraphics[width=\columnwidth]{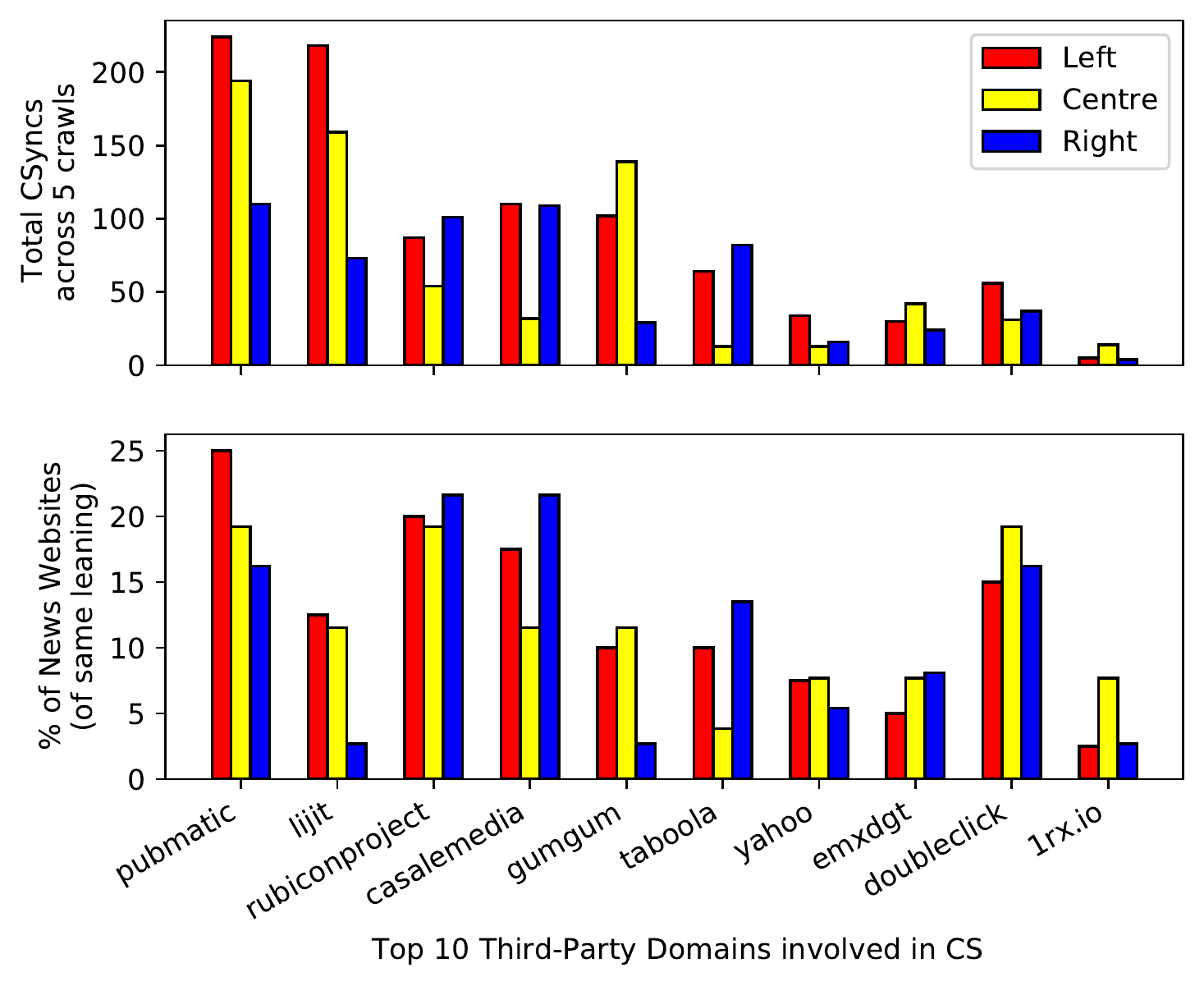}
\caption{Top 10 TPs involved in CSs, grouped by political leaning. Total CSs (top y-axis) is (TP-TP)+(TP-FP) CSs.}
 \label{fig:Top-10-TPs-in-CS}
\end{figure}

To further investigate the trackers involved in CS, we look at the domains and observe that $\sim$24\% of FPs and $\sim$18\% of TPs are performing CS.
In fact, we observe tracking domains like \textit{pubmatic.com}, which sync with other domains as high as 87 IDs.
Additionally, some IDs are synced with multiple domains.
For example, ID \textit{c3514a4b-11de-4cce-b428-365a3f6294b1-tuct65bc2e7} was found to be synced across 24 different tracking domains (from approx. 600+ TPs in our data).
Moreover, a higher median number of TPs are performing CS in Left and Centre websites than Right.
We also plot the top 10 TPs most involved in CS in Figure~\ref{fig:Top-10-TPs-in-CS}.
We observe that the top cookie-setting domains are also present here in CS.
In fact, \textit{pubmatic.com} which is setting most cookies, is also performing most CS and in most websites: $\sim$25\% Left, $\sim$19\% Centre, $\sim$16\% Right.
Also, \textit{rubiconproject.com} and \textit{doubleclick.net} perform CS in 15-22\% of websites.

\noindent \textbf{Takeaways:}
Detected user IDs are synchronized two to six times, on average, between one to five parties, on average, depending on the type of pair entity involved (TP-TP or FP-TP).
Same top domains setting cookies, appear to do heavy CS as well, covering up to 25\% of websites.
Left-leaning websites and their TPs do more CS than Right- or Centre-leaning ones.

\subsection{Device Fingerprinting}
\label{subsec:fingerprinting-results}

In this section, we present results of different fingerprinting techniques like Canvas, WebRTC, and AudioContext fingerprinting based on the methodology discussed in Sec.~\ref{subsec:fingerprinting-analysis}.
Overall, we find 32 distinct fingerprinting scripts set by 18 domains on 25.7\% of Left-, 23.7\% of Right-, and 17.9\% of Centre-leaning news websites.
Also, the most dominant type of fingerprinting is Canvas.
In particular, 26 canvas scripts are found on 23 (18.7\%) websites, from 13 unique domains; top three: \textit{jsc.mgid.com}, \textit{s0.2mdn.net}, and \textit{razorpay.com}.
Also, we find one WebRTC script set by \textit{adsafeprotected.com}, and four audioContext scripts in four websites.

\noindent \textbf{Takeaways:}
Overall, 18-25\% of FPs and TPs perform tracking using user device fingerprinting, with Left and Right adopting equally this tracking technology.

\subsection{Invisible Pixels}
\label{subsec:pixel-results}

\begin{figure}
\centering
  \includegraphics[width=0.9\columnwidth, height=5cm]{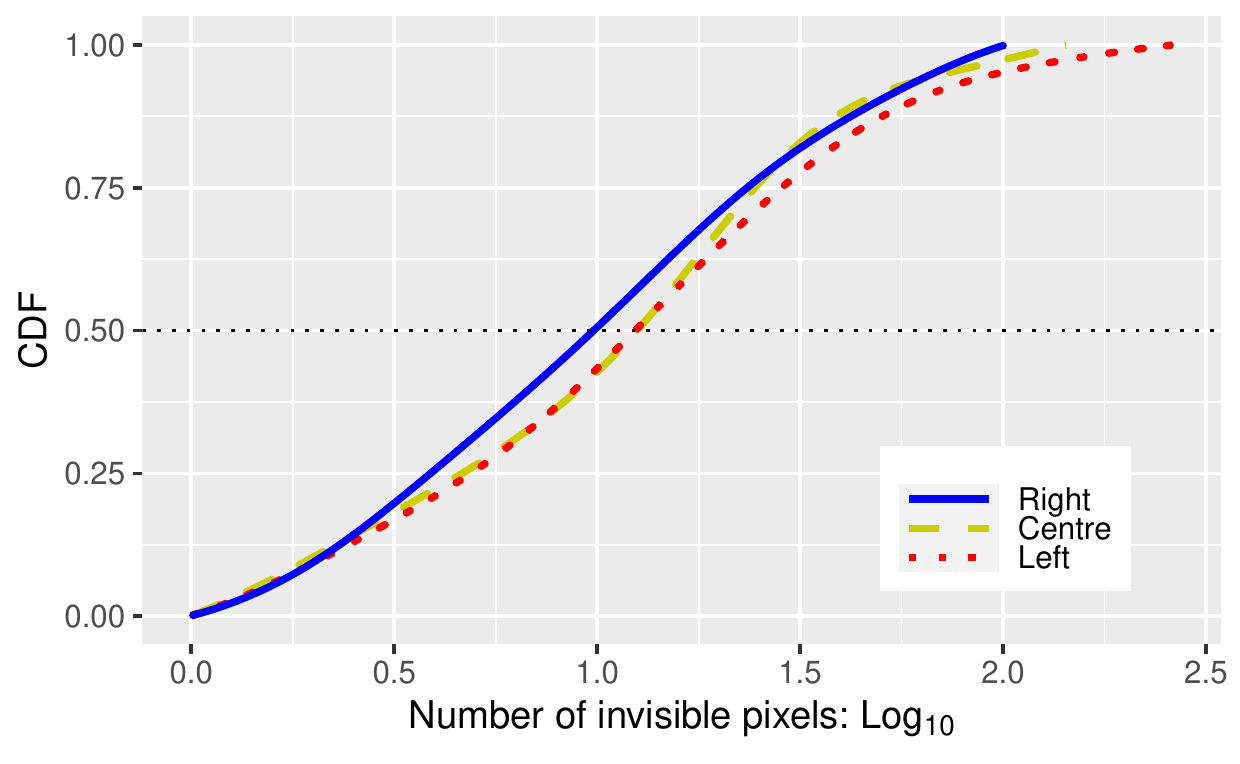}
  \caption{CDF of median number of invisible pixels for Left, Centre, and Right-leaning websites.}
  \label{fig:pixelCDF}
\end{figure}

We find 11582 images on the website homepages, out of which 5121 images have less than 1 KB size.
Following the process outlined in Sec.~\ref{subsec:pixel-analysis}, we identify 2513 invisible (1x1) pixel images, i.e., 21.7\% of all images found. Figure~\ref{fig:pixelCDF} shows the CDF of median number of invisible pixels embedded in Left-, Right-, and Centre-leaning websites.
These medians are 12, 10, and 15, respectively.
The CDF shows more intense pixel tracking by Left and Centre, than Right.

\begin{figure}
\centering
  \includegraphics[width=\columnwidth]{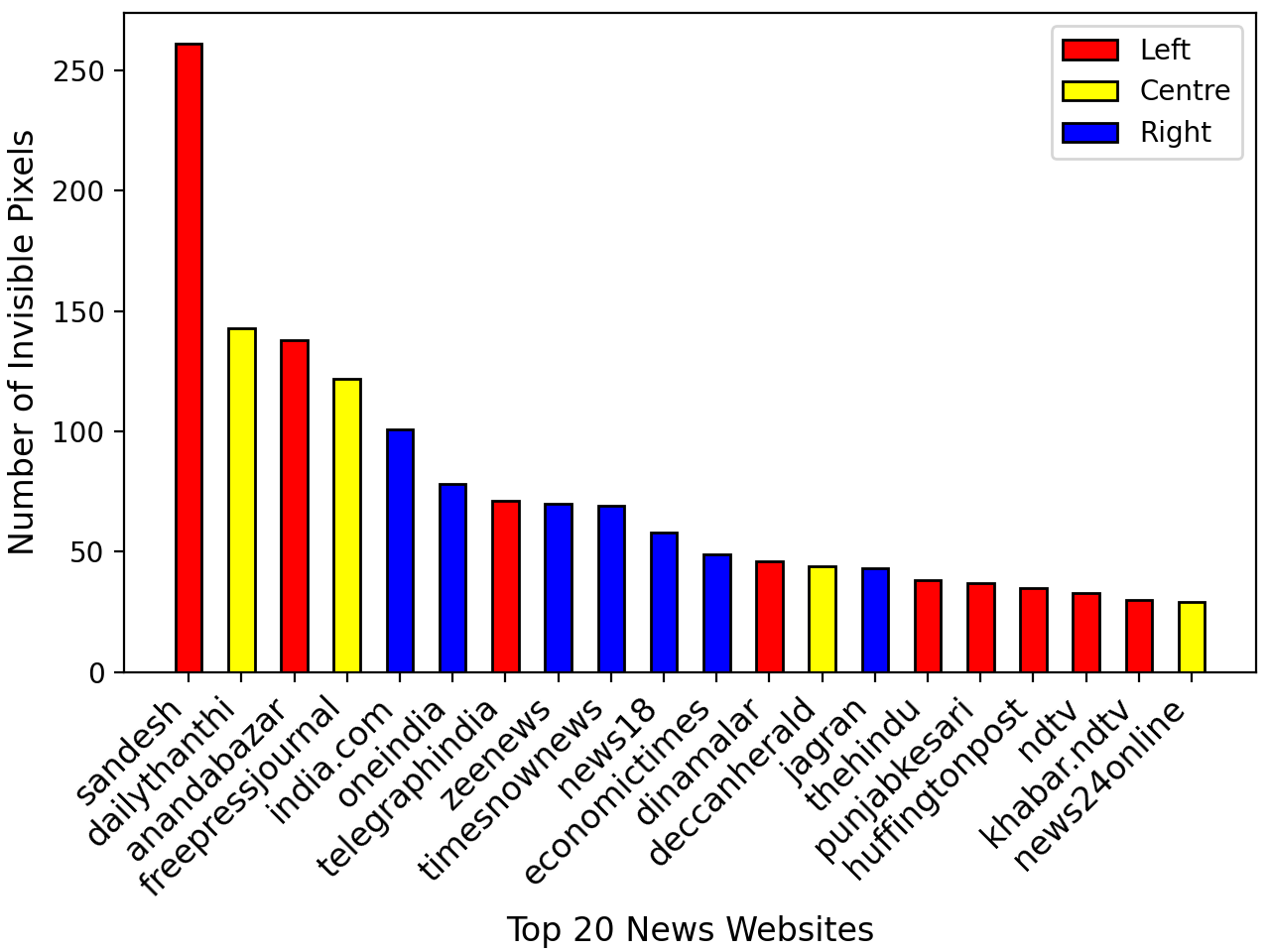}
  \caption{Top 20 news websites having invisible pixels vs. their political leanings.}
  \label{fig:inv_px_top20_websites}
\end{figure}

Figure~\ref{fig:inv_px_top20_websites} represents the top 20 FP websites having the highest number of invisible pixels, ordered by number of pixels found on their homepages.
Out of the top 20, nine are Left, seven are Right, and four are Centre.
Again, \textit{Sandesh.com} with its third-parties, earlier found to set most cookies, has the highest number of detected invisible pixels (261).
Moreover, 138 distinct TPs are detected setting these 2,513 invisible pixels.

\begin{figure}
\centering
  \includegraphics[width=\columnwidth]{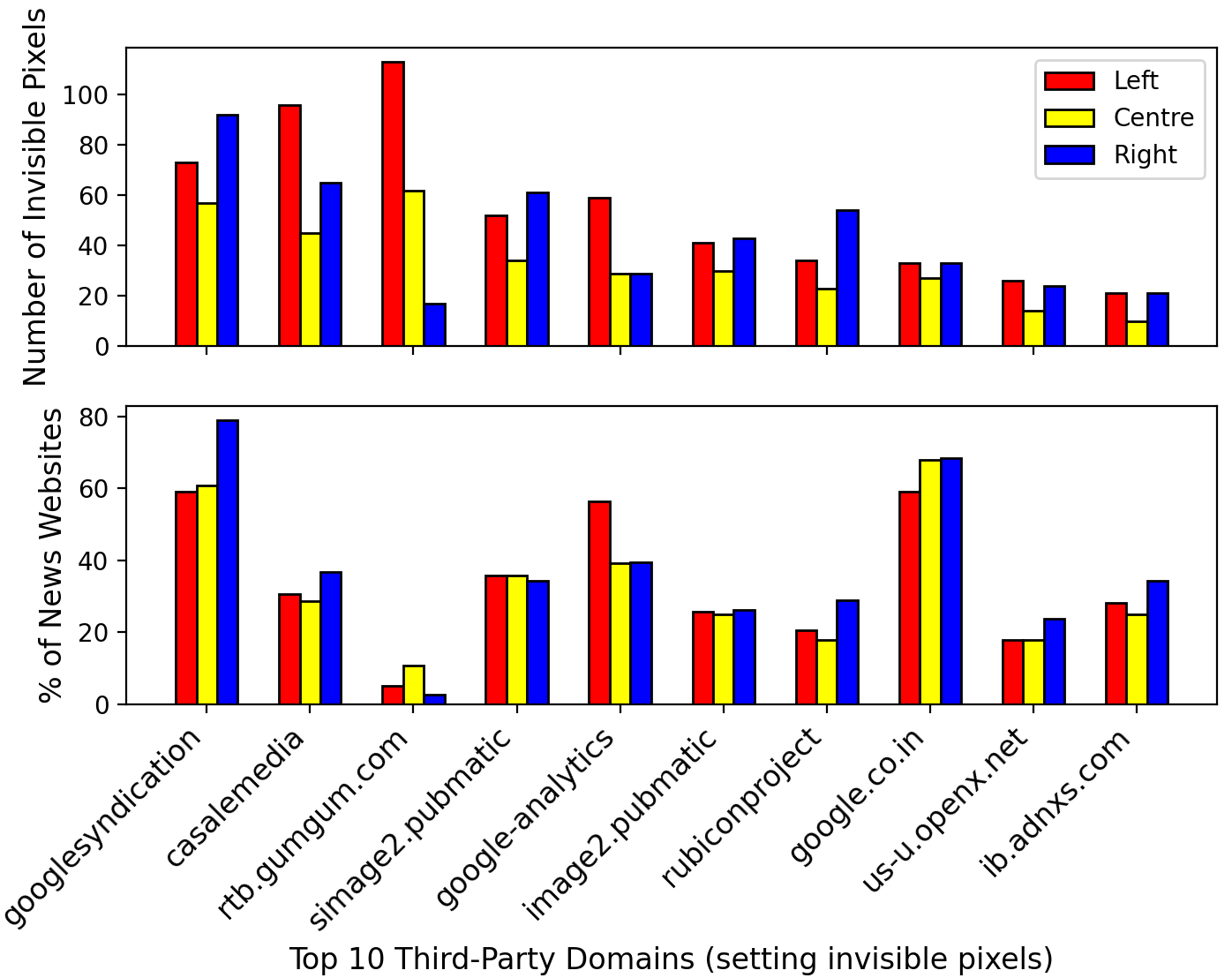}
  \caption{Top 10 Third-Party Domains setting invisible pixels on first-parties.
  Upper figure: total number of pixels set.
  Bottom figure: \% of websites embedding each third-party.}
  \label{fig:inv_px_tp}
\end{figure}

Figure~\ref{fig:inv_px_tp} shows the top 10 TPs setting invisible pixels, ordered by total number of pixels set in the news websites.
It also shows the total number of pixels set per TP.
Google-related properties (\textit{googlesyndication.com}, \textit{google-analytics.com}, and \textit{google.co.in}) dominate the market,
as the largest cumulative third-party domain that uses invisible pixels to track users' behavior on these websites.
Interesting outliers exist such as \textit{rtb.gumgum.com} that sets 113 invisible pixels on just two Left websites.

\noindent \textbf{Takeaways:}
Websites embed TPs performing invisible pixel-based tracking, with Centre-leaning websites tracking 50\% more intensely than Right, and 25\% more than Left.
Top TPs in other tracking methods (cookies, CS etc.) also perform heavy pixel-tracking, with \textit{Google} properties covering 60-80\% of the websites.

\section{Discussion \& Future Work}
\label{Sec:Discussion}

In this work, and for the first time in literature, we have done an extensive, data-driven study on the Indian online news ecosystem with respect to tracking by websites of mainstream news media with partisan leanings. The sample of news media studied have comparable resources and reach.

\noindent \textbf{Dataset:}
One of our contributions from this study is the labeled dataset of 103 news websites (reaching 77\% of Indian population) with their political leanings (Left, Right, and Centre), which we make publicly available to the research community (along with all crawls and coded methods). The aim of this paper is to show the types and extent of tracking done by mainstream news websites, which sets the essential foundation for future studies on the purpose of such targeting. Further, our findings on tracking in mobile and desktop versions is crucial as more and more Indians have started to consume news on mobile versions.

\noindent \textbf{Findings on user tracking:}
Our study shows the extensive presence of cookies irrespective of a news website's partisan leanings: on average, over 100 cookies are placed by first (FP) and third parties (TP) when visiting any of the news media websites we studied.
In general, more cookies are placed in the desktop than the mobile platforms.
Right-leaning websites place 1.2x and 1.4x the number of cookies than Centre- and Left-leaning ones in the mobile platform, whereas in the case of desktop, it is the opposite: Left tracks more than Centre and Right.
We also find that 68\% of TPs exist in both mobile and desktop versions, allowing them to perform in-depth monitoring by linking users across multiple devices.
When analyzing the categories of TPs, we find that Right- and Centre-leaning websites embed more advertising and fingerprinting TPs than Left-leaning ones.
Also, the top TP \textit{doubleclick.net} is present in 86\% of FP news websites, showing the capability of one TP domain to dominate the tracking culture across all partisan news websites in India.
Tracking with cookies goes beyond their mere presence on the browser.
About one-fourth of FPs and one-fifth of TPs are involved in cookie synchronization (CS).
We detect user IDs being synchronized close to six times (on average) between up to five parties, on average, depending on the type of syncing pair entity (TP-TP or FP-TP).
We find that the Left-leaning websites and their TPs do more CS than Right- or Centre-leaning ones.
Although around 20\% of all websites use canvas fingerprinting for tracking purposes, there is little difference between Right and Left (Centre is somewhat less) here.
In terms of invisible pixel-based tracking, TP domains in Centre-leaning websites track more than Left and the Left more than the Right.
We note that the same top TPs in other tracking methods (cookies, CS etc.) are also at the top here: \textit{Google} properties cover 60-80\% of websites, underlining the domination of the tracking market by one entity.

\noindent \textbf{Absence of Privacy Laws: ``The Wild Tracking East''.}
Our results on user tracking demonstrate that in the absence of explicit privacy laws in India, partisan websites employ different, and at times invasive tracking strategies to profile their visitors.
Left-leaning websites set more cookies, do more CS, and more pixel-based tracking, and Left and Right are almost equally intense in terms of device fingerprinting.
But what is interesting is the domination of just a few TPs that track across the studied news websites irrespective of their partisanship.
With a reach of 77\% of population from these 103 websites, the data tracked by one or few TP domains across partisan websites means that not only news websites, but even a handful of TP domains can play a very crucial role by serving political and other targeted ads.

\noindent \textbf{Implications for Privacy:}
In India, if structured privacy laws are to come into effect, online user privacy must be given high importance.
Methods of tracking currently in place can not only expose a user’s website visits and browsing histories to the tracker, but also help tracking domains to aggregate the user’s browsing patterns and interests.
These can be used to generate in-depth, detailed profiles via data synchronization through separate channels, which in turn can be exploited in numerous ways beyond just showing targeted ads.
In fact, the differential tracking across websites of different political leanings, and the opportunities offered by the above mechanics, can allow propagation of user profiles to a large number of trackers over the time.
Therefore, there is scope for these profiles being used by vested groups for targeting a user and invading the user's privacy, with the potential to influence the users visiting news websites.

\noindent \textbf{Future Work:}
The limitations of our present study along the following main lines can be tackled in future works:

\noindent \emph{1. Vernacular diversity:} Our dataset was primarily focused on websites using English language (76/103 English, with 14/103 in Hindi and 13/103 in regional languages).
Multilingual online users consist of a large portion in India~\cite{agarwal2020characterising}.
However, the diversity of languages in this country (apart from Hindi and English, India has 22 scheduled languages and several state-based official languages) raises the question: Do different political leanings perform different type and intensity of tracking across languages and news websites representing them in the regional Indian space?

\noindent \emph{2. Wide \& Complex Political Spectrum:}
Templates derived from the reference points and cases in Western settings can only partially explain the underlying political dynamics in India.
Political parties in India typically defy linear binaries of Left and Right.
In such a context, the coverage bias and media effects are variable and are contingent upon subject, personalities, and circumstances.
While the categorizations herein of ``Left'' and ``Right'' have been used as a heuristic tool, future research should dive into the contextual specifics of Indian political lines, and offer analysis with finer granularity of the political spectrum.

\noindent \emph{3. Fake News \& Hyper-partisanship:}
Recent rise in misinformation from online, hyper-partisan news websites serving fake news, coupled with tracking of users for better profiling and political ad delivery, erodes user trust in the online news ecosystem.
It requires an in-depth study of the hyper-partisan Indian news websites to assess how political websites violate their visitors' privacy.

\begin{acks}
\noindent N. Kourtellis has been partially supported by the European Union’s Horizon 2020 Research and Innovation Programme under grant agreements No 830927 (Concordia), No 871793 (Accordion), and No 871370 (Pimcity).
% These results reflect only the authors' view and the EU Commission is not responsible for any use that may be made of the information it contains.
S. Set is a Marie-Sklodowska Curie (Global India ETN) Research Fellow at King's College London. Authors also thank Arpan Gupta for helping in crawling. These results reflect only the authors' findings and do not represent the views of their institutes/organisations.
\end{acks}

\bibliography{sample-base}
\bibliographystyle{ACM-Reference-Format}

\end{document}